# Research on evolution and early warning model of network public opinion based on online Latent Dirichlet distribution model and BP neural network

Qiaozhi Bao[1*]

Department of Statistics, North Carolina State University, North Carolina, Raleigh, 27695, USA, qbao30@gatech.edu

Yanlin Chen[2]

Tandon School of Engineering, New York University, New York, Brooklyn, 11201, USA, yc3156@nyu.edu

Xusheng Ji[3]

College of Graduate and Professional Studies, Trine University, Indiana, Angola, 46703, USA, xji241@my.trine.edu

**Abstract**

Online public opinion is increasingly becoming a significant factor affecting the stability of the internet and society, particularly as the frequency of online public opinion crises has risen in recent years. Enhancing the capability for early warning of online public opinion crises is urgent. The most effective approach is to identify potential crises in their early stages and implement corresponding management measures. This study establishes a preliminary indicator system for online public opinion early warning, based on the principles of indicator system construction and the characteristics and evolution patterns of online public opinion. Subsequently, data-driven methodologies were employed to collect and preprocess public opinion indicator data. Utilizing grey relational analysis and the K-Means clustering algorithm, we classified online public opinion events into three levels: slight, warning, and severe. Furthermore, we constructed an online topic evolution model using the online Hierarchical Dirichlet Process model to analyze the thematic changes of online public opinion events across different warning levels. Finally, we developed an online public opinion early warning model using a Backpropagation (BP) neural network. The test results of early warning samples show that the model achieves high accuracy. Thus, in practical early warning applications, the BP neural network can be effectively utilized for predicting online public opinion events.

**CCS CONCEPTS**

Software and its engineering ~ Software organization and properties ~ Software system models

## 1  INTRODUCTION

With the rapid development of the internet, it has become the primary channel for individuals to express opinions, convey emotions, and obtain information. The dissemination of information on the internet is characterized by its efficiency and decentralization[Vysotskyi and Vysotskyi 2023]. Especially with the advent of new media technologies, the forms of information dissemination have undergone fundamental changes, providing internet users with entirely new avenues to express their viewpoints and opinions. These viewpoints and opinions often carry biases and subjective nuances, frequently containing rich information. When such viewpoints and opinions gain significant influence, online public opinion emerges. The volume of online public opinion data is increasing at a geometric rate, making it challenging for traditional statistical methods to analyze such vast data sets. Additionally, the internet's virtuality and openness allow users to easily share matters that pique their interest or provoke outrage. If, during this process, some users propagate false viewpoints, it can lead to misinformation and misguided public sentiment. Therefore, analyzing and establishing early warning systems for online public opinion is beneficial for timely implementation of corresponding public measures, thus maintaining social stability and government credibility[Obi et al. 2024].

## 2 RELATED RESEARCH

In the contemporary digital era, the nexus between online public opinion and security has grown increasingly tight[Xu et al. 2024]. The analysis of online public opinion security can be dependent on cloud computing storage platforms and processed through artificial intelligence algorithms to guarantee stability and security. In the research methods for online public opinion early warning, researchers initially primarily utilized traditional models such as the Delphi method, Fuzzy comprehensive judgment method, and Analytic Hierarchy Process (AHP). For instance, Luo et al. [2023]employed the Fuzzy Delphi method and FAHP to first identify the indicators for online public opinion early warning, subsequently calculating weights to establish a specific indicator system for online public opinion alerts. Qiu and Xie [2022]developed an online public opinion early warning model using the Fuzzy comprehensive judgment method, integrating AHP and authoritative weighting calculations to achieve classification of online public opinion. With the development of the internet, machine learning algorithms have been applied in the study of online public opinion warnings. Sun and Lei [2021] established an early warning model for online public opinion risk based on simulations using a BP neural network. Hu [2022]created an online public opinion early warning model based on a combination of BP neural networks and Adaboost. Zhang et al. [2021]established an evaluation method for online public opinion based on an analysis of the Bayesian model.

## 3 MODEL AND METHOD

### 3.1 Construction of early warning index system of network public opinion

The online public opinion early warning indicator system is categorized into four primary indicators: Sentiment Heat $B_1$, Sentiment Intensity $B_2$, Sentiment Direction $B_3$, and Sentiment Subject $B_4$. Each of these primary indicators is further detailed into a total of 23 secondary indicators, as illustrated in Table 1.

Table 1 Design table of primary and secondary indicators

| Primary Indicator | Secondary Indicator | Primary Indicator | Secondary Indicator |
|---|---|---|---|
| B1 Sentiment Heat | C1: Number of Images | B2 Sentiment Intensity | C7: Number of Weibo Posts |
| | C2: Number of Videos | | C8: Volume of Original Weibo Posts |
| | C3: Online Search Volume | | C9: Volume of Weibo Retweets |
| | C4: Online Discussion Volume | | C10: Volume of Weibo Comments |
| | C5: Media Coverage Volume | | C11: Volume of Weibo Likes |
| | C6: Intuitiveness of Sentiment Content | | C12: Proportion of Discussion Field |
| | | | C13: Participation of Opinion Leaders |
| B3 | C14: Rate of Change in Online Search Volume | B4 | C20: Blogger Identity |

| Primary Indicator | Secondary Indicator | Primary Indicator | Secondary Indicator |
| --- | --- | --- | --- |
| Sentiment Direction | C15: Rate of Change in Online Discussion Volume | Sentiment Subject | C21: Number of Verified Weibo Accounts |
| | C16: Rate of Change in Original Weibo Posts Volume | | C22: Number of Followers |
| | C17: Rate of Change in Weibo Retweets Volume | | C23: Blogger Influence |
| | C18: Rate of Change in Weibo Comments Volume | | |
| | C19: Rate of Change in Weibo Likes Volume | | |

### 3.2 Classification and analysis of early warning levels of network public opinion

#### 3.2.1 Grey Relation Analysis

Given the characteristics of uncertainty and suddenness inherent in online public opinion, Grey Relation Analysis can be employed to grade its impact levels. First, the maximum value from the online public opinion early warning indicators is selected as the reference sequence, while the data from each time period serves as the comparative sequence. Next, by calculating the relation factors and degrees of association between the comparative sequence and the reference sequence at each time point, the warning level of public opinion events at each time can be analyzed. The specific calculation process is as follows:

First, determine the reference sequence and comparative sequences for the risk level assessment of online public opinion events. Reference Sequence $X_0 = (x_0(1), x_0(2), \ldots, x_0(n))$, Comparative Sequences $X_m = (x_m(1), x_m(2), \ldots, x_m(n))$. Due to the different units of various indicators, direct comparison is not feasible. Therefore, before using Grey Relation Analysis for comparison, it is necessary to standardize the indicators. Next, compute the association factors between the reference sequence $X_0$ and the comparative sequence for each indicator. The association factor $\eta_i(k)$ between the comparative sequence $X_i (i = 1,2,\ldots,m)$ and the reference sequence $X_0$ is defined as follows:

$$\eta_i(k) = \frac{\min_i \min_k |X_0(k) - X_i(k)| + \rho \max_i \max_k |X_0(k) - X_i(k)|}{|X_0(k) - X_i(k)| + \rho \max_i \max_k |X_0(k) - X_i(k)|} \quad (1)$$

where $\rho \in (0,1)$ is the resolution factor. A smaller $\rho$ indicates greater discriminative power, with $\rho$ generally taken as 0.5. Next, calculate the degree of association between the reference sequence and the comparative sequence. Since the association factor $\eta_i(k)$ represents the values of the association factors between the reference sequence and the comparative sequence across components and the data volume is large, it is not easy to compare directly. Thus, it is necessary to aggregate the association factors of the various indicators of sub-public opinion events at each moment into a single value, termed the degree of association. The degree of association $\gamma_i$ between the comparative sequence $X_i (i = 1,2,\ldots,m)$ and the reference sequence $X_0$ is defined as $\gamma_i = \frac{1}{n}\sum_{k=1}^{n} \eta_i(k)$. The degrees of association are then arranged in order of magnitude, constituting an association sequence, denoted as $X$. The higher the degree of association, the higher the aggregate score. Specifically, if $\gamma_i > \gamma_j$, it can be stated that $\{X_i\}$ is superior to $\{X_j\}$ regarding the same parent public opinion event $\{X_0\}$, denoted as $\{X_i\} > \{X_j\}$.

*3.2.2 K-Means clustering algorithm*

The article employs the K-Means clustering algorithm to classify the warning levels of online public opinion and uses the degree of association between various sub-public opinion events and the primary public opinion event as the basis for classification and the allocation of warning levels. Let us assume a given sample of public opinion event data $X$, which contains $n$ objects with $m$-dimensional feature indicator values represented as $X = \{X_1, X_2, X_3, \ldots, X_n\}$. The objective of the K-Means clustering algorithm is to group the $n$ objects into $k$ specific classes based on their similarities, where each object belongs to only one class. The main steps of the K-Means clustering algorithm are as follows: 1. Initialize Cluster Centers Initialize $k$ cluster centers $\{C_1, C_2, C_3, \ldots, C_k\}$, where $1 < k \leq n$. 2. Calculate the distance from each object to the $k$ cluster centers using Euclidean distance $dis(X_i, C_j) = \sqrt{\sum_{t=1}^{m}(X_{it} - C_{jt})^2}$, Here, $X_i$ represents the $i$-th object, $1 \leq i \leq n$; $C_j$ denotes the $j$-th cluster center, $1 \leq j \leq k$; $X_{it}$ represents the $t$-th feature indicator value of the $i$-th object; and $C_{jt}$ denotes the $t$-th feature indicator value of the $j$-th cluster center, $1 \leq t \leq m$. 3.Clustering. Calculate the distance from each object to each cluster center, and assign each object to the class of the nearest cluster center, resulting in $k$ classes $\{S_1, S_2, S_3, \ldots, S_k\}$. 4. Calculate Mean. Compute the mean of all objects in each class and update it as the new cluster center $C_l = \frac{\sum_{X_i \in S_l} X_i}{|S_l|}$. Here, $C_l$ denotes the $l$-th cluster center, $1 \leq l \leq k$; $|S_l|$ represents the number of objects in the $l$-th class; $X_i$ denotes the $i$-th object in the $l$-th class, $1 \leq i \leq |S_l|$. 5. Iterative Update. Continuously iterate to assign objects and update cluster centers until there is no change in the cluster centers of all classes or a specific number of iterations has been reached. If not, return to step (2).

**3.3 BP neural network model**

In the online public opinion early warning indicator system established in this paper, a total of 10 warning indicators are included; therefore, the number of input nodes is set to 10. Since the BP neural network algorithm possesses a large number of parallel distributed structures and nonlinear dynamic characteristics, the number of nodes in the hidden layer is determined by the following formula, based on a review of relevant literature and materials $N = \sqrt{m + n} + a$, where $m$ is the number of input nodes, $n$ is the number of output nodes, and $a$ is any constant between 1 and 10. In this study, six different scenarios were tested with the number of hidden layer nodes set to 5,6,7,8,9, and 10. It was found that with 6 hidden layer nodes, the algorithm achieved good accuracy, resulting in a final configuration of 6 hidden layer nodes. During the processing in the hidden layer, the BP neural network functions as a "black box." The training parameters used in this study are detailed in Table 2.

Table 2 Training parameter setting

| Parameter | Value | Parameter | Value |
| --- | --- | --- | --- |
| Input - Hidden Layer Transfer Function | tansig | Number of Training Iterations | 1000 |
| Hidden - Output Layer Transfer Function | purelin | Learning Rate | 0.01 |
| Training Function | trainlm | Minimum Training Target Error | 0.00001 |

## 4 CASE STUDY AND EVALUATION

### 4.1 Source of dataset

As the Internet matures and expands, the outbreak of online public opinion crises has become increasingly frequent. On March 21, 2022, an Eastern Airlines Boeing 737-800 passenger plane crashed. The incident occurred suddenly, resulting in significant casualties and losses, sparking extensive comments and debates from netizens and various sectors of society regarding the rescue efforts and the causes of the accident. This event, characterized by its suddenness and high public interest, garnered significant attention and widespread dissemination among netizens, making it a highly representative case. Therefore, this paper selects the "March 21 Eastern Airlines Plane Crash" as a sample for case collection. This study collected data over 36 days, from March 18, 2022, to April 22, 2022, dividing the time into 36 time slices (days). The samples from these 36 time slices were used to construct a training group for the model and a testing group to verify the prediction accuracy of the constructed model. The first 25 time slices of sample data were designated as the training group, while the remaining 11 time slices constituted the testing group.

### 4.2 Correlation analysis

To determine the correlation coefficients among various indicators and reduce the redundancy of the indicator system, a correlation analysis was first conducted on the standardized data. This paper calculated the Pearson correlation coefficient between each pair of indicators to assess their degree of correlation, as indicated in the following formula:

$$r_{ab} = \frac{s_{ab}^2}{s_a s_b} = \frac{\sum_{i=1}^{n}(a_i-\bar{a})(b_i-\bar{b})}{\sqrt{\sum_{i=1}^{n}(a_i-\bar{a})^2 \sum_{i=1}^{n}(b_i-\bar{b})^2}} \quad (2)$$

where $r_{ab}$ is the correlation coefficient between indicators $a$ and $b$, $a_i$ is the $i$-th value of indicator $a$, $b_i$ is the $i$-th value of indicator $b$, $\bar{a}$ is the mean of indicator $a$, $\bar{b}$ is the mean of indicator $b$, and $n$ is the sample size. The correlation coefficients between each set of indicators were compared with a set significance threshold $M$. If $|r_{ab}| \geq M$, it indicates a very significant correlation between the two indicators, allowing for the deletion of one of them. Conversely, if $|r_{ab}| < M$, it indicates that the relationship between the two indicators is not significant, and both indicators can be retained. In this study, $M$ was set to 0.85. The primary objective of principal component analysis is to reduce the number of variables while ensuring minimal information loss. The model is described by the following equation $F_j = a_{i1}X_1 + a_{i2}X_3 + \cdots + a_{im}X_m, i,j = 1,2,\cdots,k$. where $X_i$ represents the indicators, $F_j$ represents the principal components, $a_{im}$ is the $m$-th component of the $i$-th feature vector, and $m$ is the number of indicators. This study retains those principal components whose cumulative variance contribution is greater than 80% and whose factor load absolute values exceed 0.8, while discarding the rest. Tables 3 and 4 present the variance contribution rates of the online public opinion early warning indicator system and the principal component factor load coefficients, respectively.

Table 3 Variance Contribution Rates of the Online Public Opinion Early Warning Indicator System

| Primary Indicator | First Principal Component Variance Contribution Rate | Cumulative Variance Contribution Rate |
|---|---|---|
| Public Sentiment Heat $B_1$ | 0.82 | 0.82 |

| Public Sentiment Intensity $B_2$ | 0.81 | 0.81 |
| Public Sentiment Tendency $B_3$ | 0.81 | 0.81 |

Table 4 Principal Component Factor Load Coefficients of the Online Public Opinion Early Warning Indicator System

| Primary Indicator | Secondary Indicator | First Principal Component | Selection Result |
| --- | --- | --- | --- |
| Public Sentiment Heat $B_1$ | Number of Images $C_1$ | 0.93 | Retained |
| | Network Search Volume $C_3$ | 0.92 | Retained |
| | Media Coverage Volume $C_5$ | 0.98 | Retained |
| | Intuition of Public Sentiment Content $C_6$ | 0.81 | Retained |
| | Number of Weibo Posts $C_7$ | 0.95 | Retained |
| Public Sentiment Intensity $B_2$ | Number of Weibo Comments $C_{10}$ | 0.90 | Retained |
| | Opinion Leader Participation $C_{13}$ | 0.83 | Retained |
| | Change Rate of Network Search Volume $C_{14}$ | 0.66 | Deleted |
| Public Sentiment Tendency $B_3$ | Change Rate of Original Weibo Posts $C_{16}$ | 0.88 | Retained |
| | Change Rate of Weibo Comments $C_{18}$ | 0.92 | Retained |

### 4.3 Soundness test

The rationality test criteria for the indicator system can be measured using the information contribution rate. The information contribution rate represents the amount of information conveyed by the indicators, which can be reflected by the standard deviation of the indicator data. By comparing the standard deviation of the final indicator data with that of the original indicator data, we can obtain the amount of information represented by the final indicator system. If the selected indicators can express 90% of the information from the initial indicators, the rationality of the indicator system can be established. The formula for calculating the information contribution rate is as follows $IN = \frac{1}{s}\sum_{i=1}^{s} \sigma_i / \frac{1}{p}\sum_{j=1}^{p} \sigma_j$. Where $\sigma_i$ is the standard deviation of the filtered indicator data, $\sigma_j$ is the standard deviation of the preliminary indicator data before filtering, $s$ is the number of indicators after filtering, and $p$ is the number of preliminary indicators before filtering. According to calculations, the information contribution rate $IN = 98.9\%$, indicating that the indicator system filtered through correlation analysis and principal component analysis can represent 98.9% of the information. The final 10 secondary indicators represent the majority of the indicators, demonstrating that the network public opinion indicator system constructed in this paper is reasonable and effective.

Based on the final network public opinion early warning indicator system, the dimensionless sample data of the standardized indicator system is obtained, followed by setting a reference sequence. Using the maximum value from each indicator, the reference sequence can be obtained:

$$X_0 = (3.0229, 3.5068, 2.4998, 2.3620, 2.7690, 4.4564, 1.9951, 3.5582, 3.1701, 2.7078) \quad (3)$$

Each set of indicator data from each time point serves as a comparison sequence, with a total of 36 comparison sequences. Gray correlation analysis is utilized to determine the correlation degree between the comparison sequences and the reference sequence.

Table 5 Sample data correlation degree

| Date | Degree of association | Date | Degree of association | Date | Degree of association | Date | Degree of association |
| --- | --- | --- | --- | --- | --- | --- | --- |
| 3.18 | 0.4817 | 3.27 | 0.6716 | 4.5 | 0.4836 | 4.14 | 0.5105 |
| 3.19 | 0.4808 | 3.28 | 0.5718 | 4.6 | 0.4766 | 4.15 | 0.4761 |
| 3.20 | 0.4813 | 3.29 | 0.4987 | 4.7 | 0.4731 | 4.16 | 0.4744 |
| 3.21 | 0.7120 | 3.30 | 0.5474 | 4.8 | 0.4737 | 4.17 | 0.5094 |
| 3.22 | 0.9200 | 3.31 | 0.5694 | 4.9 | 0.4726 | 4.18 | 0.4895 |
| 3.23 | 0.6767 | 4.1 | 0.4869 | 4.10 | 0.4708 | 4.19 | 0.4724 |
| 3.24 | 0.6729 | 4.2 | 0.5021 | 4.11 | 0.5694 | 4.20 | 0.5274 |
| 3.25 | 0.7001 | 4.3 | 0.4993 | 4.12 | 0.4933 | 4.21 | 0.4904 |
| 3.26 | 0.6306 | 4.4 | 0.4864 | 4.13 | 0.4794 | 4.22 | 0.4648 |

As can be seen from the table above, the correlation degree suddenly increased on March 21, reaching its maximum value on March 22. This indicates that the occurrence of the event attracted significant attention from netizens. The speed of online public opinion dissemination is rapid at this time, necessitating timely monitoring and guidance.

Using the K-Means clustering algorithm to determine the warning level classification thresholds for public opinion events, the 36 calculated correlation degrees are clustered using the K-Means algorithm. The results yield three clustering centers: the first clustering center is 0.4845, the second clustering center is 0.5693, and the third clustering center is 0.7256. Thus, the warning levels of events in various time periods can be categorized into three classes—minor level, warning level, and severe level—based on the values of their correlation degrees, using the K-Means clustering algorithm. After obtaining the three clustering centers through the K-Means algorithm, the distance of each case from the public sentiment event during different time periods to the clustering centers is used to derive the value ranges for the minor level, warning level, and severe level. Specifically, the correlation degree range for the minor level is [0.00, 0.52], the correlation degree range for the warning level is [0.52, 0.64], and the correlation degree range for the severe level is [0.64, 1.00]. We made the network public opinion early warning system according to the above algorithm, and the functions of the system are shown in the following figure 1.

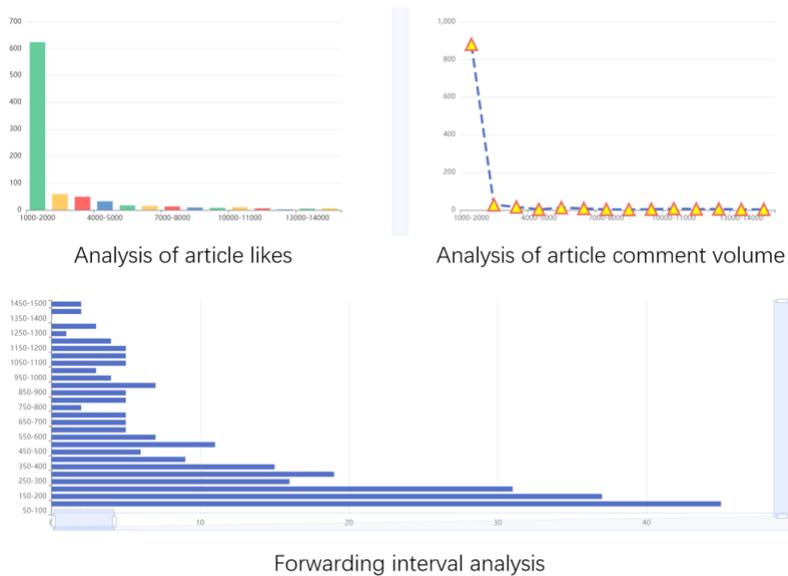

Figure 1 Network public opinion early warning system part of the function diagram

### 4.4 Model verification

First, each time slice of the original data was set as a separate time node. Since there were a total of 36 time slices, 36 time nodes were established. Among them, the first 25 time nodes were used as the training set, while the remaining 11 time nodes were designated as the test set. After standardizing the data, experiments were conducted using MATLAB. The training and test datasets were input in matrix form, with the expected output variables determined based on the warning levels: mild level was set to 1, alert level to 2, and severe level to 3. Figure 2 shows the expected output for training and testing samples.

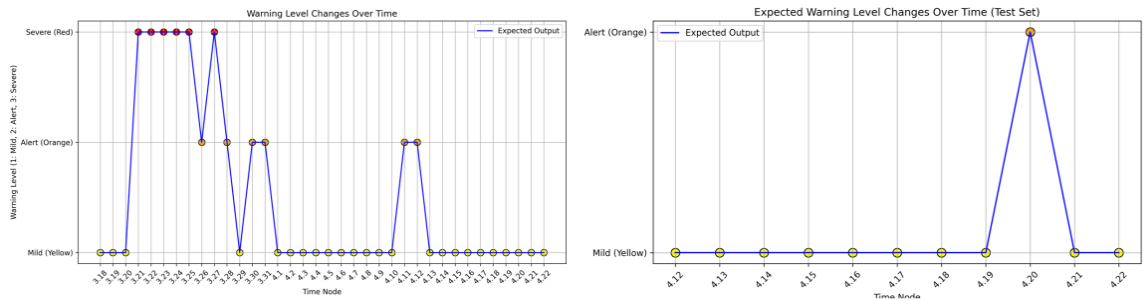

Figure 2 The expected output of the sample

Through the training of the model, when the error rate is $1.7352e{-}6$, the target range is reached, the running results are shown in Table 6, and the training error curve change diagram is shown in Figure 3.

Table 6 BP Neural Network Operation Results

| Progress | Epoch | Performance | Gradient | Validation Checks |
|---|---|---|---|---|
| Value | 6 | 1.7352e-6 | 3.21e | 0 |

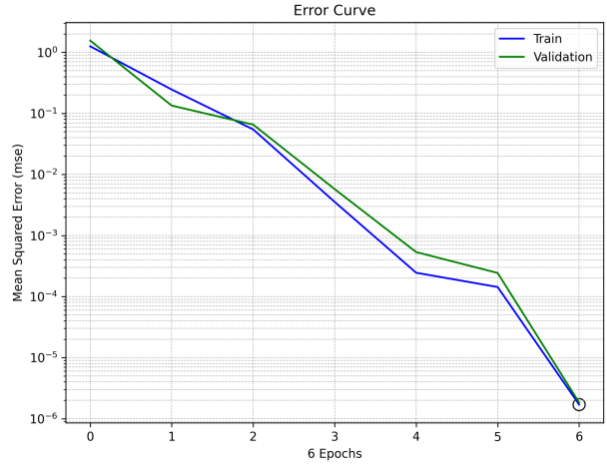

Figure 3 Plot of training error curve change

By inputting the samples of the standardized test group and dividing the early warning levels of different samples by the public opinion early warning model, the early warning accuracy of the BP neural network early warning model is judged. Table 7 shows the sample data of the test group.

Table 7 Test group sample data

| Time node | $C_1$ | $C_3$ | $C_5$ | $C_6$ | $C_7$ | $C_{10}$ | $C_{13}$ | $C_{16}$ | $C_{18}$ | $C_{23}$ |
|---|---|---|---|---|---|---|---|---|---|---|
| 4.12 | -0.84 | -0.94 | -0.85 | -0.43 | -0.79 | -0.98 | 0.19 | -0.70 | -0.23 | -0.80 |
| 4.13 | -0.91 | -0.95 | -0.97 | -0.39 | -0.96 | -1.00 | -0.58 | -0.48 | -0.01 | -0.97 |
| 4.14 | -0.97 | -0.96 | -0.99 | 0.76 | -0.90 | -0.99 | -0.99 | -0.30 | 0.01 | -0.95 |
| 4.15 | -0.97 | -0.97 | -0.99 | -0.43 | -0.96 | -1.00 | -0.90 | -0.31 | 0.00 | -0.98 |
| 4.16 | -0.97 | -0.97 | -0.99 | -0.43 | -0.98 | -1.00 | -0.98 | -0.32 | 0.00 | -0.99 |
| 4.17 | -0.90 | -0.97 | -0.99 | -0.77 | -0.81 | -0.90 | 0.40 | -0.11 | 0.09 | -0.81 |
| 4.18 | -0.93 | -0.96 | -0.97 | -0.53 | -0.90 | -0.99 | 0.01 | -0.38 | -0.08 | -0.91 |
| 4.19 | -0.99 | -0.97 | -1.00 | -0.91 | -0.97 | -1.00 | -0.52 | -0.38 | -0.01 | -0.98 |
| 4.20 | -0.77 | -0.93 | -0.90 | -0.75 | -0.58 | -0.78 | 0.25 | 0.15 | 0.21 | -0.59 |

| 4.21 | -0.86 | -0.96 | -0.91 | -0.21 | -0.81 | -0.99 | -0.11 | -0.60 | -0.20 | -0.84 |
| 4.22 | -0.99 | -0.98 | -0.98 | -1.04 | -0.97 | -1.00 | -0.85 | -0.46 | -0.01 | -0.98 |

The error between the predicted value and the expected value of only one sample is very large, and then the prediction results of the test group samples are transformed into the warning level $Y_i$. The prediction results of the test group samples are shown in Figure 4 below:

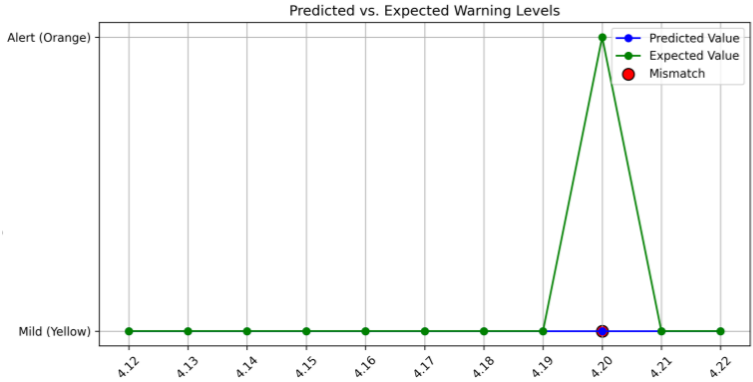

Figure 4 Prediction results for the test group sample

As can be seen from Figure 4, the prediction of only one sample among the 11 test groups is different from the expectation, and the prediction results of the other samples are all correct. Therefore, the network public opinion early warning model based on BP neural network can be obtained, and its early warning accuracy is 90.91%. Therefore, the use of BP neural network for early warning of network public opinion is effective.

### 4.5  Explain and use BP neural network prediction

With the widespread application of artificial intelligence technology in social science research and public policy formulation, effectively interpreting and utilizing the predictive results of deep learning models, including Back Propagation (BP) neural networks, has become a critical research topic. For decision-makers, models are not merely "black boxes"; the logic behind their predictions and the key influencing factors are equally important. Understanding these logics and factors not only enhances the precision of policy interventions but also improves the traceability and interpretability of decisions. For small-scale BP neural networks, weight matrix visualization can provide a rough understanding of the importance of connections between inputs and outputs. However, for deeper or larger-scale networks, relying solely on weight visualization often fails to offer clear policy implications. Additionally, weight magnitudes must be interpreted in conjunction with the scales of different input variables to prevent misinterpretations caused by differences in measurement units. By computing the partial derivatives of the output with respect to input variables, one can observe the sensitivity of each input variable to the predictive results. A higher sensitivity of a particular input variable indicates a greater impact on the model's output.

For decision-makers, this means prioritizing factors that have the greatest influence on public opinion when designing policy interventions or allocating resources. One approach to evaluating the importance of a specific feature is to randomly shuffle its values while keeping other features unchanged and then measure the change in model performance. Since this method relies on re-evaluating model performance, it effectively reflects the contribution of individual variables to predictive accuracy. The Local Interpretable Model-Agnostic Explanations (LIME) method generates new samples around a specific prediction point and fits them using a simple, interpretable model (such as a linear model) to obtain a locally interpretable linear approximation. In the context of public opinion prediction, LIME helps decision-makers identify which input features contribute most to a given prediction instance (e.g., public opinion data from a particular region). By using LIME, policymakers can understand, at a micro level, why certain groups receive higher (or lower) predictions and further trace the attitudinal characteristics of these groups toward relevant social issues.

## 5 CONCLUSION

Currently, online public opinion has become a major aspect influencing societal sentiment. Therefore, in the context of sudden public opinion crises, researching corresponding online public opinion warning models is a necessary measure. The themes of online public opinion are closely related to the warning levels of online public opinion. The analysis of the evolution of theme words in public opinion events using the OLDA topic model reveals that as the intensity of public opinion heat increases—particularly in the severe level (red)—the focus of netizens becomes more concentrated, and the topics of discussion are more focused. The application of the BP neural network in the online public opinion warning model has yielded good results. The warning model constructed was tested using a sample set from the test group, and the results indicate that the warning accuracy of the online public opinion warning model using the BP neural network is 90.91%, demonstrating a significantly good performance.

**Use of AI**

I used chatgpt4o to touch up the whole text. After using these tools, I reviewed and edited the content as needed and take full responsibility for the content of the publication. Since I am a non-native English speaker, I used it for writing assistance to improve grammar and text expression.